\documentclass[final,3p,times,twocolumn]{elsarticle}

\usepackage{setspace}
\usepackage{amssymb}
\usepackage{amsmath}
\usepackage{bm}

\usepackage{siunitx}
\usepackage{setspace}
\usepackage{float}
\usepackage{graphicx}
\usepackage{placeins}

\usepackage{xcolor}

\usepackage{lineno}

\makeatletter
\def\ps@pprintTitle{%
   \let\@oddhead\@empty
   \let\@evenhead\@empty
   \let\@oddfoot\@empty
   \let\@evenfoot\@empty}
\makeatother

\journal{}

\newcommand{\pp}[1]{\left(#1\right)}
\newcommand{\norm}[1]{\left|#1\right|}

\newcommand{\vm}[1]{\bm{#1}}

\newcommand{\dealii}[0]{\texttt{deal.II}}
\newcommand{\lethe}[0]{\texttt{Lethe}}

\newcommand*\justify{%
  \fontdimen2\font=0.4em
  \fontdimen3\font=0.2em
  \fontdimen4\font=0.1em
  \fontdimen7\font=0.1em
  \hyphenchar\font=`\-
}

\renewcommand{\texttt}[1]{%
  \begingroup
  \ttfamily
  \begingroup\lccode`~=`/\lowercase{\endgroup\def~}{/\discretionary{}{}{}}%
  \begingroup\lccode`~=`[\lowercase{\endgroup\def~}{[\discretionary{}{}{}}%
  \begingroup\lccode`~=`.\lowercase{\endgroup\def~}{.\discretionary{}{}{}}%
  \catcode`/=\active\catcode`[=\active\catcode`.=\active
  \justify\scantokens{#1\noexpand}%
  \endgroup
}

\begin{document}

\newcommand{\overlap}{\delta_{\mathrm{n}}}
\newcommand{\tdisplacement}{\vm{\delta_{\mathrm{t}}}}
\newcommand{\me}{m_{\mathrm{e}}}
\newcommand{\muT}{\mu_{\mathrm{t}}}
\newcommand{\muTE}{\mu_{\mathrm{t,e}}}
\newcommand{\muR}{\mu_{\mathrm{r}}}
\newcommand{\muRE}{\mu_{\mathrm{r,e}}}
\newcommand{\Ee}{e_{\mathrm{e}}}
\newcommand{\Ree}{R_\mathrm{e}}
\newcommand{\CRD}[1]{$\mathrm{CRD}_{#1}$}
\newcommand{\LRD}[1]{$\mathrm{LRD}_{#1}$}
\begin{frontmatter}

\title{Increase in packing density during multi-layer powder spreading: An experimental and numerical study.}

\author[1,2]{Olivier Gaboriault}
\author[3]{Anatolie Timercan}
\author[3]{Roger Pelletier}
\author[3]{Louis-Philippe Lefebvre}
\author[2]{David Melancon} 
\author[1]{Bruno Blais\footnote{Corresponding author.
Email address: bruno.blais@polymtl.ca (Bruno Blais)}}

\affiliation[1]{organization={Department of Chemical Engineering, High-performance Automatization Optimization and Simulation (CHAOS) laboratory, Polytechnique Montréal},
            addressline={2500 Chemin de Polytechnique}, 
            city={Montréal},
            postcode={H3T 1J4}, 
            state={Québec},
            country={Canada}}
\affiliation[2]{organization={Department of Mechanical Engineering, Laboratory for Multiscale Mechanics (LM2), Polytechnique Montréal},
            addressline={2500 Chemin de Polytechnique}, 
            city={Montréal},
            postcode={H3T 1J4}, 
            state={Québec},
            country={Canada}}
\affiliation[3]{organization={National Research Council Canada},
            addressline={75 Bd de Mortagne}, 
            city={Boucherville},
            postcode={J4B 6Y4}, 
            state={Québec},
            country={Canada}}

\begin{graphicalabstract}
\includegraphics[width=\textwidth]{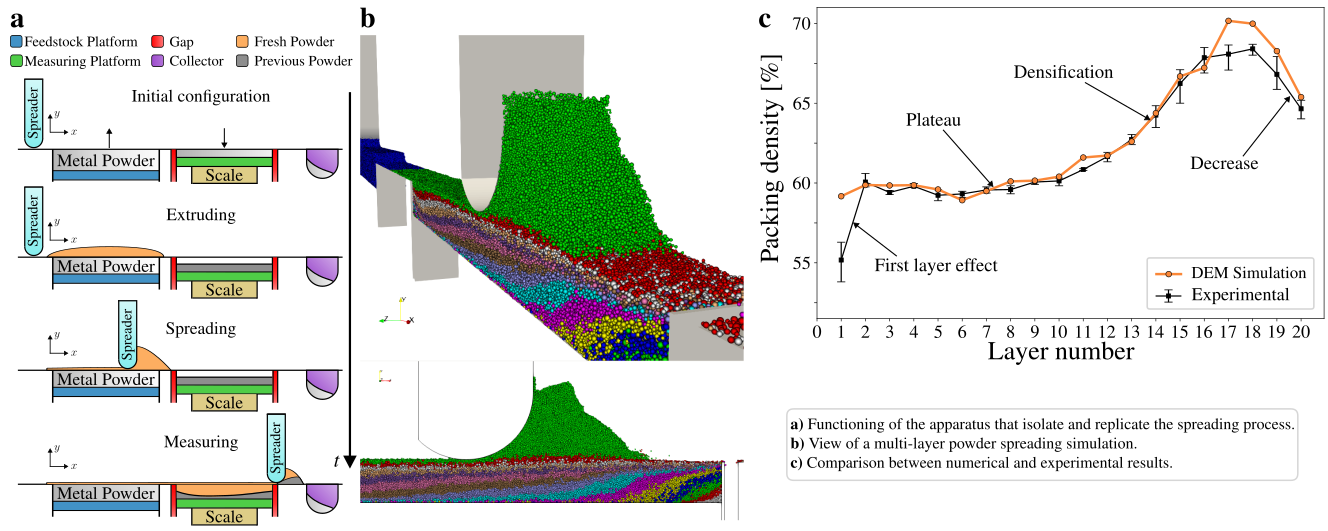}
\end{graphicalabstract}

\begin{highlights}
    \item First report of the existence of an increase in effective layer packing density occurring at higher layer number (5 to 10).
    \item Large-scale multi-layer powder spreading DEM simulations for metal additive manufacturing. 
    \item The DEM results show that a significant amount of powder is removed from the powder bed during the spreading of subsequent layers.
\end{highlights}

\begin{keyword}
Multi-layer powder spreading, Discrete element method, Additive manufacturing, Powder deposition mechanism
\end{keyword}

\begin{abstract}
A custom apparatus designed to isolate and replicate the spreading process of metal powder in additive manufacturing demonstrates a sudden and unexplained increase in packing density beyond layers 5 to 10. We replicate the experiments that lead to densification with the discrete element method (DEM) using \lethe{}, an open-source software framework. We show that large-scale multi-layer DEM simulations are able to reproduce the densification observed experimentally. Using the Lagrangian simulation results, we highlight significant particle displacement in the powder bed at lower layer number, accompanied by static zones generated by the vertical wall surrounding the powder bed. The amplitude of the densification and the layer number at which it starts to occur is correlated to the distance between those two vertical walls which delimit the powder spreading area. This study addresses the gap between mono-layer powder spreading studies on hard-flat surfaces and the actual metal powder-based additive manufacturing processes by providing a better understanding of how the powder bed behaves during multi-layer spreading.
\end{abstract}

\end{frontmatter} 

\section{Introduction}
\label{sec:introduction}
In the context of metal additive manufacturing, the powder spreading process consists of spreading thin layers of metal powder on a building plate, which are then selectively melted, or sintered, using an energy source. This process of spreading then melting is repeated layer-by-layer up until the final product is built. Inefficient powder spreading leads to low packing density, poor surface uniformity and poor homogeneity of the powder layers. This negatively affects the melting process \cite{yap_2015} and causes defects such as porosity, balling or lack of fusion in the final part \cite{khairallah_2016,mostafaei_2022} which may lead to process failure. As a result, the powder spreading step is critical to the success of the additive manufacturing process, but ensuring high-quality powder layers can decrease the efficiency of the manufacturing process by increasing the build time due to a reduced spreading velocity \cite{shaheen_2021}. Therefore, the influence of the process parameters (e.g. spreading velocity, layer thickness, recoater geometry) on the layer quality should be assessed to create a robust and fast spreading step. This assessment can be difficult to perform experimentally due to the microscale at which the interaction between the powder and the recoater occurs, which is why many researchers have studied it using numerical simulations.

With the exception of Jaggannagari et al. \cite{jaggannagari_2025} and a few other works investigating spreading over consolidated layers \cite{wang_2022a, long_2024}, most numerical studies have focused on the spreading of a single layer of powder on a hard and flat surface \cite{chen_2017a, chen_2019, chen_2020, haeri_2017a, meier_2019a, he_2020} which is not representative of the entire powder spreading process. In addition, most multi-layer studies lack experimental comparison or fail to replicate the experimental conditions adequately, for instance by employing a different spreader geometry in the experiments and in the simulations~\cite{jaggannagari_2025}.

\begin{figure}[!htpb]
        \centering
        \includegraphics[width=75mm]{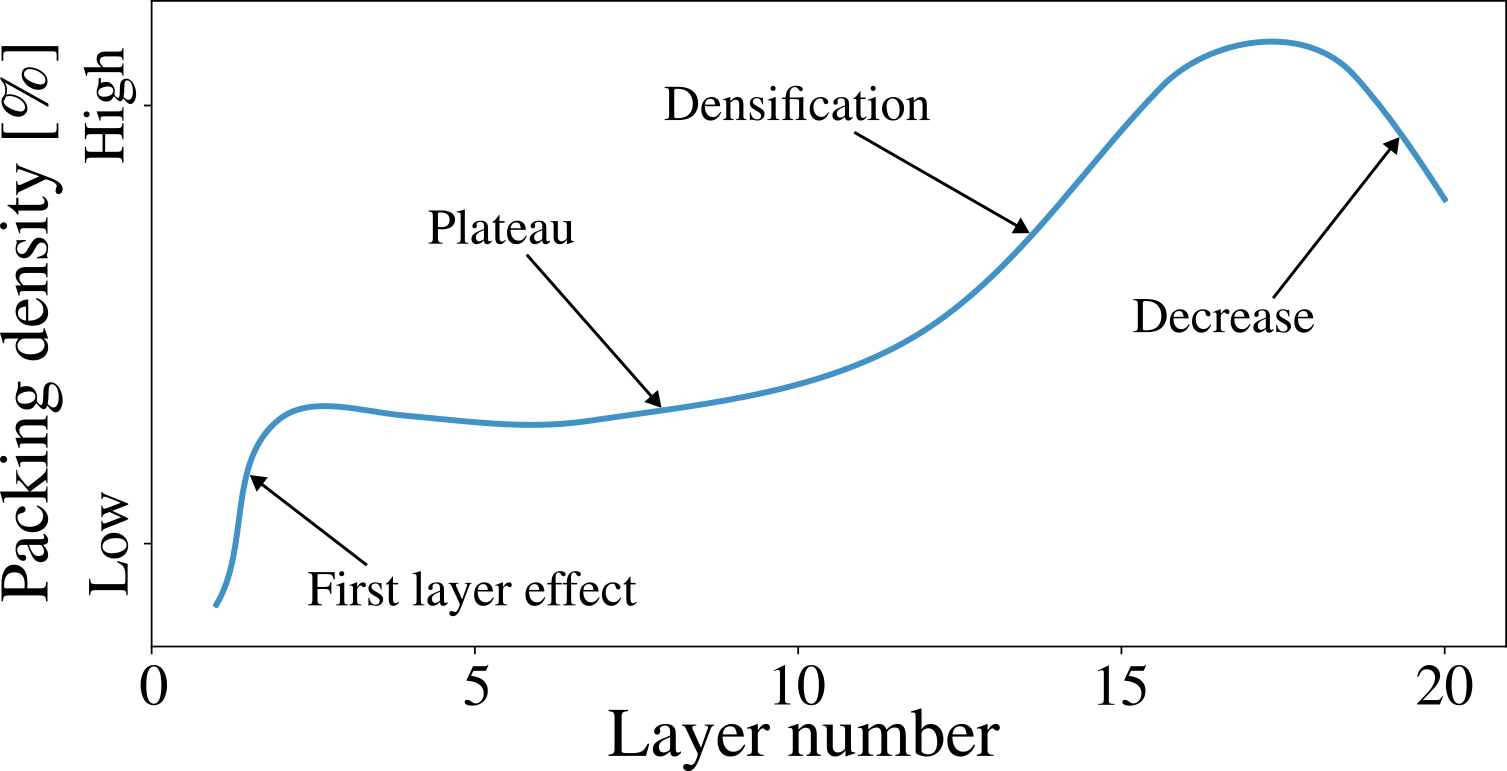}
        \caption{\textbf{Schematic of the increase in packing density observed over many experiments.} Packing density as a function of layer number showing an initial increase due to the first layer effect followed by plateau region, densification and decrease during the 20 first layers.}
        \label{fig::exp_result_intro}
\end{figure}

To address the gap surrounding numerical multi-layer powder spreading studies with experimental comparison, a custom experimental apparatus was designed and built by the National Research Council of Canada. This device measures the effective packing density of consecutive layers after each spreading pass. Figure~\ref{fig::exp_result_intro} presents a conceptual representation of the packing density measurements obtained with this apparatus as a function of the layer number. The measurements exhibit four distinct regimes: first-layer effect, plateau, densification, and decrease. These regimes are repeatable across different types of powders (e.g., titanium, aluminium, stainless steel) and reveal an unexplained increase in packing density at higher layer numbers during the densification regime, which, to the authors’ knowledge, has not been documented in previous studies. This densification suggests that the initial layers could achieve higher packing density under different sets of process parameters. A better understanding of this phenomenon could guide the development of improved spreading strategies by minimizing its impact on the powder bed and by increasing the homogeneity of the spreading process over multiple layers.

To investigate the source of this densification phenomenon, we turn to numerics and perform large-scale, multi-layer discrete element method (DEM) simulations which replicate the experiments. First, we detail in Section~\ref{sec::experimental} the custom experimental apparatus used to measure the powder packing density. In Section~\ref{sec::num_meth}, we present the DEM framework as well as the methodology used to simulate the powder spreading experiments. We then show in Section~\ref{sec::results} that the DEM model is able to reproduce the densification observed experimentally. We find that the densification is amplified when the length of the build plate is increased. Using the particle displacement between the consecutive layers, we give insights into the physics that drive this densification phenomenon before concluding on the potential outcome of this work for multi-layer powder spreading.

\section{Experimental methodology}
\label{sec::experimental}

\subsection{Experimental apparatus}
\label{subsec::experimental_apparatus}
\begin{figure}[!htbp]
        \centering
        \includegraphics[width=75mm]{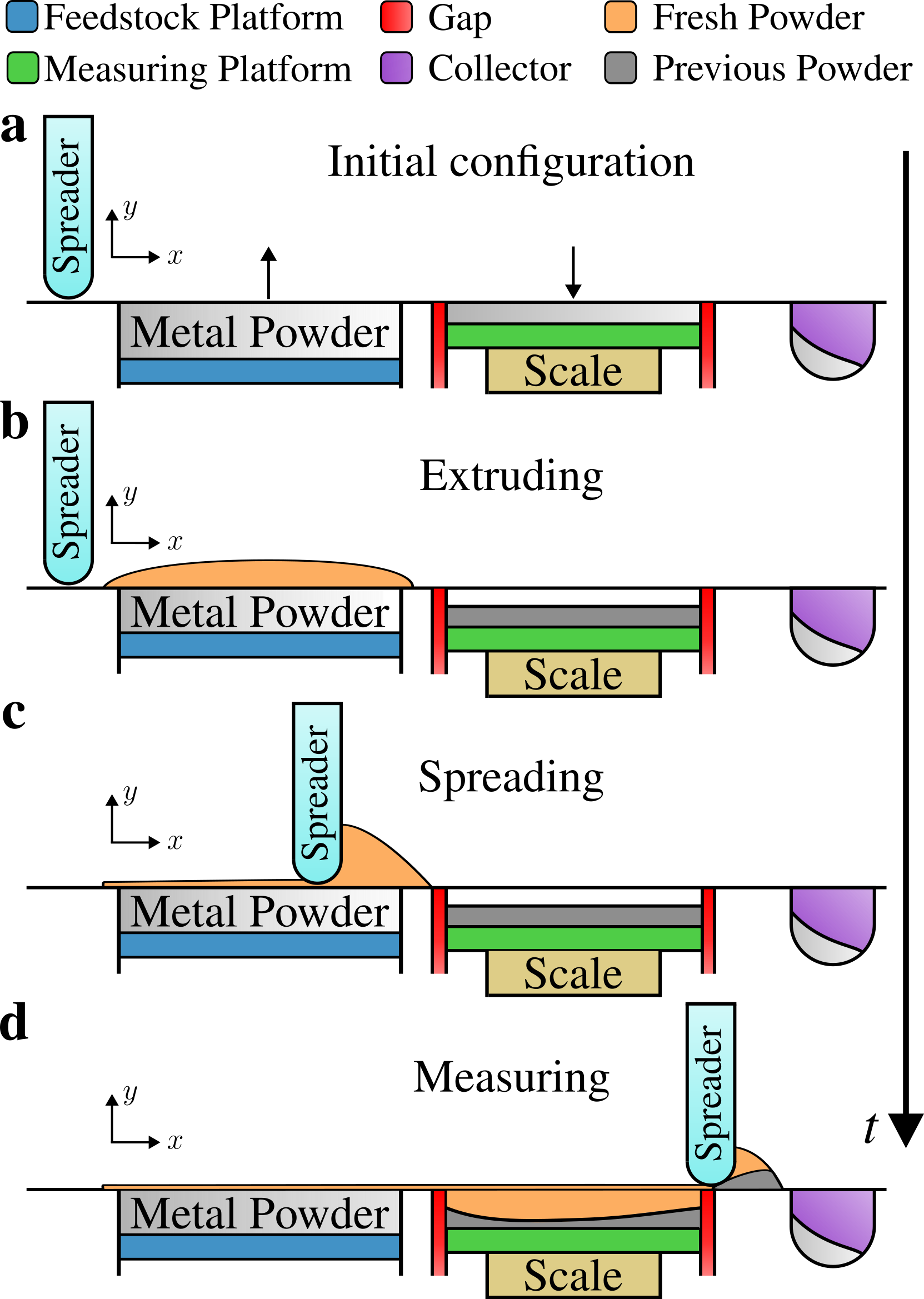}
        \caption{\textbf{Functioning of the apparatus that isolate and replicate the spreading process.} Side view schematic of the experimental apparatus showing each step of the spreading process. \textbf{(a)} Initial configuration, \textbf{(b)} Extrusion, \textbf{(c)} Spreading, and \textbf{(d)} Measuring steps.}
        \label{fig::exp_set_up_side}
\end{figure}

\begin{figure}[!htpb]
        \centering
        \includegraphics[width=\linewidth]{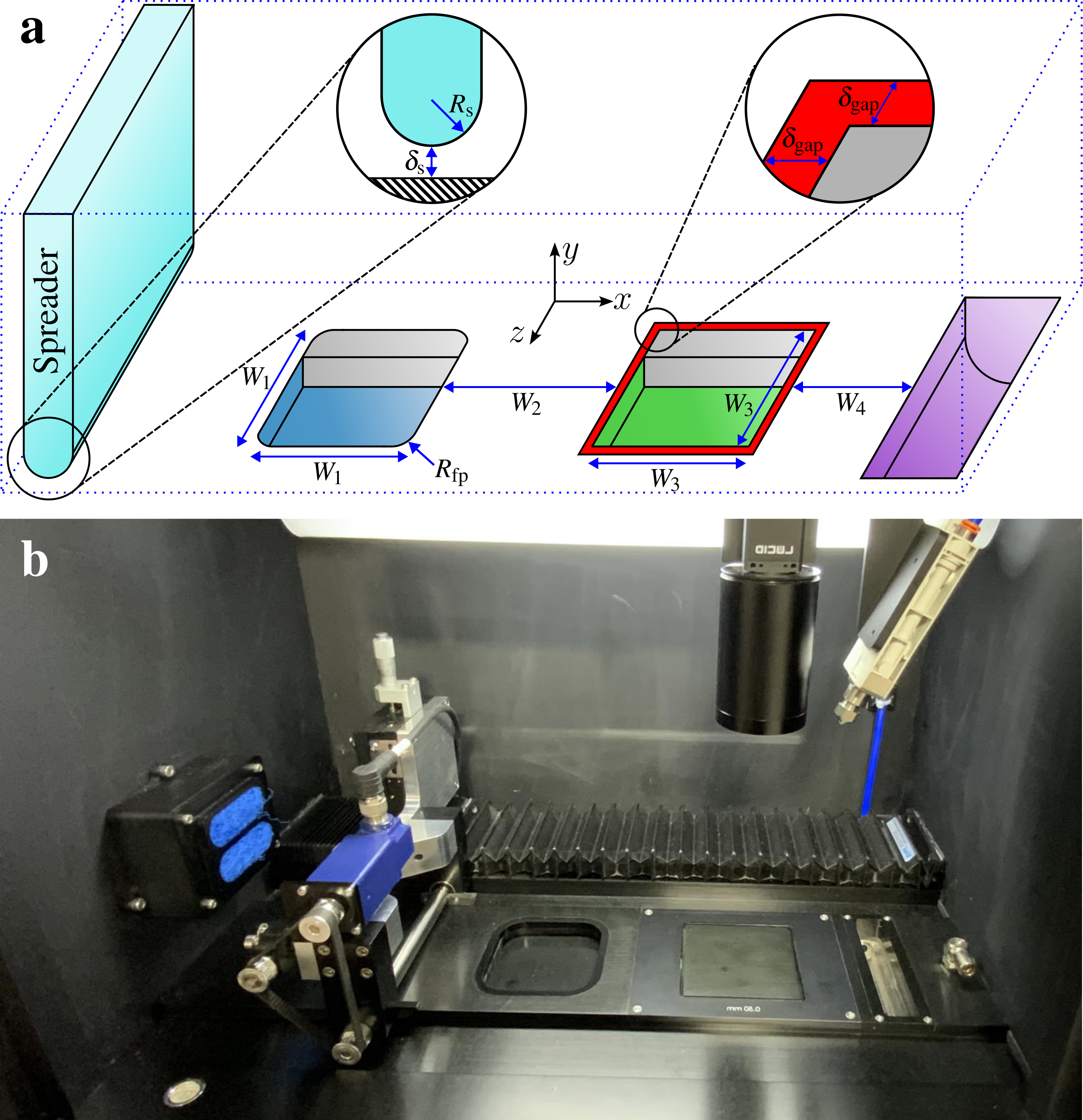}
        \caption{\textbf{Overview of the apparatus} \textbf{(a)} Isometric schematic view of the experimental apparatus showing important dimension variables of the equipment. \textbf{(b)} Photo of the experiment equipment.}
        \label{fig::exp_set_up_iso}
\end{figure}

The custom apparatus measures the packing density of the entire powder bed after each deposited layer. Figure \ref{fig::exp_set_up_side} illustrates the sequence used for each measurement, starting with the initial configuration (Figure \ref{fig::exp_set_up_side}a). Fresh powder is extruded from the feedstock platform while the measuring plate is lowered using Micronix ES-50PM-11500 piezoelectric motors (Figure \ref{fig::exp_set_up_side}b). A spreader then distributes the fresh powder over the measuring plate by moving at constant velocity (Figure \ref{fig::exp_set_up_side}c). Finally, the excess powder is retrieved in the collector. A Sartorius WZA1203-NC scale with a resolution of $\SI{1}{\milli\gram}$ is placed beneath the measuring plate to register the mass of powder deposited after each spread layer. The measuring plate is surrounded by a $\SI{500}{\micro\meter}$ gap ensuring that only the powder over the measuring plate is weighed (Figure~\ref{fig::exp_set_up_side}d). This fully automated apparatus, visible in Figure~\ref{fig::exp_set_up_iso}b, can use interchangeable spreader geometry, variable spreading velocity ($v_s$) from 2 to 200 \si{\milli\meter\per\second}, feedstock platform vertical displacement ($\delta_f$) from 0.05 to 32 \si{\milli\meter} and adjustable layer thickness ($\delta_{l}$) from 20 to 200 \si{\micro\meter}. Although this apparatus allows for wide range of operating parameters, we limit this study to the single set listed in Table \ref{tab::exp_metho::op_param}.

\begin{table}[!htpb]
\caption{Operating parameters used for experiments}
\centering
    \begin{tabular}{l | l}
        Operating parameters           & Value \\   
        \hline
        Layer thickness ($\delta_{l}$) & $\SI{100}{\micro\meter}$ \\
        Spreading velocity ($v_s$)     & $\SI{100}{\milli\meter\per\second}$\\
        Feedstock platform displacement ($\delta_f$) & $600$-$\SI{2200}{\micro\meter}$ \\
        \label{tab::exp_metho::op_param}
    \end{tabular}
\end{table}

Figure~\ref{fig::exp_set_up_iso}a presents an isometric schematic view of the spreading apparatus, highlighting the key dimensions relevant to the study, which are listed in Table \ref{tab::exp_app_dimensions}. $W_1$ and $W_3$ are both the length and depth of the feedstock platform and measuring platform, respectively. $W_2$ is the length of the transfer plate between both platforms and $W_4$ is the distance between the measuring platform and the collector. $R_\mathrm{s}$ is the spreader radius and $\delta_\mathrm{s}$ is the vertical distance (in $y$) between the tip of the spreader and the transfer plate. Finally, $R_\mathrm{fp}$ is the radius of the feedstock platform corners and $\delta_\mathrm{gap}$ is the gap size. 

\begin{table}[!htpb]
\caption{Relevant dimensions of the apparatus}
\centering
\begin{tabular}{l | l}
    Variable     & Value \\           
    \hline
    $W_1$ & $\SI{69}{\milli\meter}$\\
    $W_2$ & $\SI{56}{\milli\meter}$\\
    $W_3$ & $\SI{73}{\milli\meter}$\\
    $W_4$ & $\SI{46.5}{\milli\meter}$\\
    $R_\mathrm{s}$ & $\SI{5}{\milli\meter}$\\ 
    $\delta_\mathrm{s}$ & $\approx \SI{0.}{\milli\meter}$\\
    $R_\mathrm{fp}$ & $\SI{16}{\milli\meter}$\\ 
    $\delta_\mathrm{gap}$ & $\SI{500}{\micro\meter}$\\
    \end{tabular}
    \label{tab::exp_app_dimensions}
\end{table}

\subsection{Powder}
\label{subsec:powder}
\begin{figure}[!htb]
        \centering
        \includegraphics[width=\linewidth]{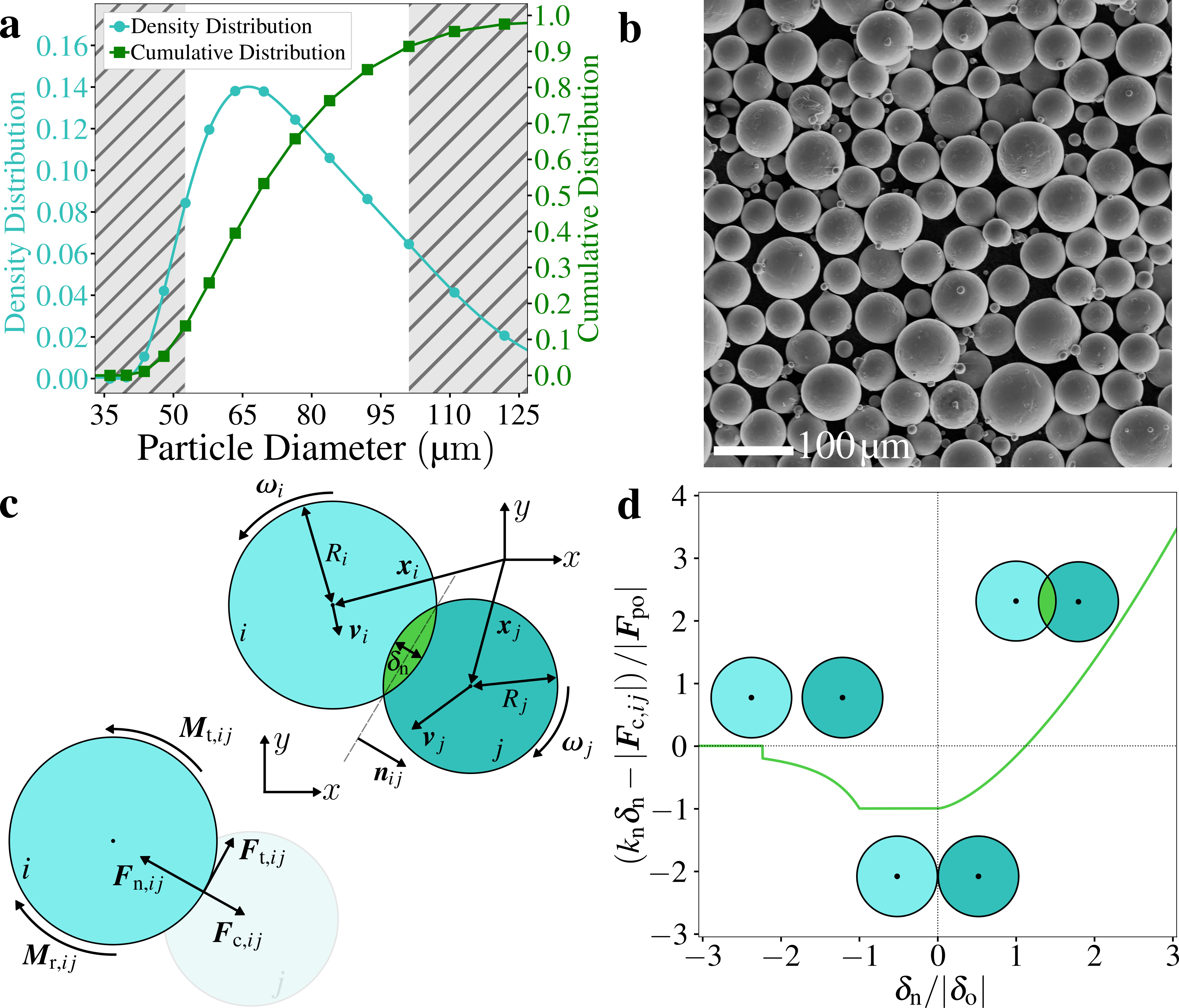}
        \caption{\textbf{Powder particle size distribution and numerical modeling} \textbf{(a)} Particle size distribution of the Ti-6Al-4V powder from AP\&C used in this study with shaded area representing the excluded size range for the simulations. \textbf{(b)} SEM images, produced with a NANOS tabletop SEM from Semplor at \SI{15}{\kilo\volt}, of the Ti-6Al-4V powder used in the experimental work. \textbf{(c)} Schematic representation of particle contact mechanics showing the interactions between particles i and j, including the normal force ($\vm{F}_{\mathrm{n},ij}$), tangential force ($\vm{F}_{\mathrm{t},ij}$), cohesive force ($\vm{F}_{\mathrm{c},ij}$) and torque due to the tangential force ($\vm{M}_{\mathrm{t},ij}$) and the rolling resistance torque ($\vm{M}_{\mathrm{r},ij}$). \textbf{(d)} Normalized force interaction plot showing total normal contact force as a function of the normalized overlap (excluding the damping term) with different contact conditions illustrated, i.e., no contact (left), zero overlap (center) and contact (right).}
        \label{fig::PSD_SEM_Model_DMT}
\end{figure}

We use commercially available Ti-6Al-4V powder made by the \textit{Advanced Powders and Coatings (AP\&C)} company for all experiments in this study. Figure~\ref{fig::PSD_SEM_Model_DMT}a presents this powder's particle size distribution (PSD). The blue curve illustrate the density distribution while the green curve the cumulative distribution, both based on mass, with shaded regions indicating excluded size ranges for the DEM simulations. The $d_{10}$ and $d_{90}$, which are the particle diameter of the 10th and 90th percentile of the PSD, are $45$ and $\SI{106}{\micro\meter}$, respectively. Figure~\ref{fig::PSD_SEM_Model_DMT}b shows a scanning electron microscope (SEM) image of this powder, made with a NANOS tabletop SEM from Semplor at 15 kV. Visually, the powder presents few satellites and the particles are spherical.

\subsection{Spreading experiment} 
\label{subsec:powder_layer}
Each experiment consists in spreading 21 layers of powder with a given set of operating parameters. The first layer (layer 0) uses a higher dosing factor corresponding to a vertical displacement of the feedstock platform of \SI{2200}{\micro\meter} to ensure full coverage of the measuring plate. Subsequent layers use a \SI{600}{\micro\meter} vertical displacement. Using the weight measured after each layer, the cumulative relative density (\CRD{}) of the powder bed and the effective relative density (\LRD{}) of the newly spread layer can be calculated using equations \eqref{eq::cumu_rel_density} and \eqref{eq::eff_rel_density}. 
\begin{align}
    \mathrm{CRD}_k &= \frac{m_{k} - m_{1}}{\rho A_{mp} \delta_{l} (k-1)} \label{eq::cumu_rel_density}, \\
    \mathrm{LRD}_k &= \frac{m_{k} - m_{k-1}}{\rho A_{mp} \delta_{l} }\label{eq::eff_rel_density}, 
\end{align}
where $m_{k}$ is the total mass of the powder on the measuring plate after the layer $k$, $\rho$ is the bulk density of the powder metal alloy, $A_{mp}$ is the surface area of the measuring plate equal to $\SI{5329}{\milli\meter\squared}$ and $\delta_{l}$ is the layer thickness. Note that \LRD{} and \CRD{} measurements start at layer $k=1$ and $k=2$, respectively.

\section{Numerical methodology}
\label{sec::num_meth}
All of our simulations are carried out using \lethe{} \cite{blais_2020, alphonius_2025}, an open-source computational fluid dynamics and DEM \cite{golshan_2023} software based on \dealii{}, a \texttt{C++} finite element library \cite{africa_2024}. 

\subsection{Discrete element method}
\label{subsec:DEM}
\lethe{}'s DEM solver employs the soft-sphere approach which allows particles to slightly overlap upon contact (See Figure \ref{fig::PSD_SEM_Model_DMT}c). The resulting overlaps are used to compute contact and cohesive forces and torques between particles. These forces and torques are then used when solving Newton's second law \eqref{eq::dem_newton} and Euler's law of angular motion \eqref{eq::dem_euler} on each particles via an explicit velocity Verlet time integration scheme.

\begin{align}
     m_{i}\frac{d \vm{v}_{i}}{dt} &= m_{i}\vm{g} + \sum_{j \in \mathrm{C}_i}  \vm{F}_{\mathrm{n},ij} + \vm{F}_{\mathrm{t},ij} + \vm{F}_{\mathrm{c},ij}  \label{eq::dem_newton},\\
    I_{i}\frac{d\vm{\omega}_{i}}{dt} &= \sum_{j \in \mathrm{C}_i} \vm{M}_{\mathrm{t},i} +  \vm{M}_{\mathrm{r},ij},
    \label{eq::dem_euler}
\end{align}
where, $i$ and $j$ are particle indices, $\mathrm{C}_i$ denotes the list of particles $j$ that are in interaction with $i$, $m_{i}$ and $I_{i}$ are the mass and moment of inertia of the particles, $\vm{v}_{i}$ and $\vm{\omega}_{i}$ are the translational and angular velocities and $\vm{g}$ is the gravitational acceleration. $\vm{F}_{\mathrm{n},ij}$, $\vm{F}_{\mathrm{t},ij}$ and $\vm{F}_{\mathrm{c},ij}$ are the normal, tangential and cohesive forces resulting from the particle interaction. Finally, $\vm{M}_{\mathrm{t},ij}$ and $\vm{M}_{\mathrm{r},ij}$ are the contact torque due to the tangential force and the rolling resistance torque, respectively. The force models to calculate these forces and torques are defined in Section \ref{subsec:Contact_force}, \ref{subsec:Cohesive_force} and \ref{subsec:rolling_res}.

\subsubsection{Contact forces and torque}
\label{subsec:Contact_force}
Following the standard approach in the literature, we use a non-linear visco-elastic model to compute the normal and tangential contact forces from the overlap, tangential displacement and relative velocity between the particles \cite{tsuji_1992}. To be concise, we only present the equations for particle–particle interaction, as the particle–wall models follow the same principle. Figure~\ref{fig::PSD_SEM_Model_DMT}c illustrates the contact between two particles, $i$ and $j$, and illustrates various quantities used in equations \eqref{eq::dem_normal_overlap} to \eqref{eq::dem_normal_velocity}:
\begin{align}
    \delta_{\mathrm{n}} &= R_i + R_j - \norm{\vm{x}_j - \vm{x}_i} \label{eq::dem_normal_overlap}  \\
    \vm{n}_{ij} &= \frac{\vm{x}_j - \vm{x}_i}{\norm{\vm{x}_j - \vm{x}_i}}\\
    \vm{v}_{ij} &= \vm{v}_j - \vm{v}_i + \pp{R_i \vm{\omega}_i + R_j \vm{\omega}_j} \times \vm{n}_{ij} \label{eq::dem_tot_rel_vel} \\ 
    \vm{v}_{\mathrm{n},ij} &=  \pp{\vm{n}_{ij} \cdot \vm{v}_{ij}} \vm{n}_{ij} \label{eq::dem_normal_velocity}
    \\\vm{v}_{\mathrm{t},ij} &=\vm{v}_{ij} - \vm{v}_{\mathrm{n},ij}.\label{eq::dem_tangential_velocity}
\end{align}
where $R_i$ is the radius. $\vm{n}_{ij}$ is the normal unit contact vector between particle $i$ and $j$ while $\vm{v}_{ij}$ is the total relative velocity at the point of contact with $\vm{v}_{\mathrm{n},ij}$ and $\vm{v}_{\mathrm{t},ij}$ being its normal and tangential component, respectively.

When $\delta_{\mathrm{n}}$ is positive, particles $i$ and $j$ are in contact, thus $\vm{F}_{\mathrm{n},ij}$, $\vm{F}_{\mathrm{t},ij}$, and $\vm{M}_{\mathrm{t},i}$ must be computed using the model coefficients listed in Table \ref{model:table:DEMForces} and with equations \eqref{eq::normal_force} to \eqref{eq::tangential_torque}: 

\begin{table*}[!htbp]
\caption{Equations for the DEM contact and cohesive force models}
\centering
\begin{tabular}{|l l|}
  \hline
    Parameter                     & Equation \\           
    \hline
    Normal stiffness              & $k_{\mathrm{n}}= \frac{4}{3}Y_{\mathrm{e}} \sqrt{R_{\mathrm{e}}\delta_{\mathrm{n}}}$ \\
    Tangential stiffness          & $k_{\mathrm{t}}= 8G_{\mathrm{e}} \sqrt{R_{\mathrm{e}} \delta_{\mathrm{n}}}$ \\
    Normal damping                & $\eta_{\mathrm{n}}= -2\sqrt{\frac{5}{6}}  \frac{\ln\pp{e_\mathrm{r}}}{\sqrt{\ln^2\pp{e_\mathrm{r}} + \pi^2}} \sqrt{\frac{2}{3}k_{\mathrm{n}} m_{\mathrm{e}}}$ \\
    Tangential damping            & $\eta_{\mathrm{t}}= -2\sqrt{\frac{5}{6}}  \frac{\ln\pp{e_\mathrm{r}}}{\sqrt{\ln^2\pp{e_\mathrm{r}} + \pi^2}} \sqrt{k_{\mathrm{t}} m_{\mathrm{e}}}$ \\
    Effective mass                & $\frac{1}{m_{\mathrm{e}}} = \frac{1}{m_i} + \frac{1}{m_j}$ \\
    Effective radius              & $\frac{1}{R_{\mathrm{e}}} = \frac{1}{R_{i}} + \frac{1}{R_j}$ \\
    Effective Young's modulus     & $\frac{1}{Y_{\mathrm{e}}} = \frac{\pp{1-\nu_i^2}}{Y_i} + \frac{\pp{1-\nu_j^2}}{Y_j}$ \\
	Effective shear modulus      & $\frac{1}{G_{\mathrm{e}}} = \frac{2\pp{2+\nu_i}\pp{1-\nu_i}}{Y_i} + \frac{2\pp{2+\nu_j}\pp{1-\nu_j}}{Y_j}$ \\
	Effective sliding friction coefficient                & $\mu_{\mathrm{t}} = \pp{\frac{1}{\mu_{\mathrm{t},i}} + \frac{1}{\mu_{\mathrm{t},j}}}^{-1}$ \\
	Effective rolling friction coefficient                &  $\mu_{r} = \pp{\frac{1}{ \mu_{r,i}}  + \frac{1}{ \mu_{r,j}}}^{-1}$\\
    Effective restitution coefficient     & $e_r = \pp{\frac{1}{ e_{r,i}}  + \frac{1}{ e_{r,j}}}^{-1}$ \\
    Effective surface energy  & $\gamma_{\mathrm{e}} = \gamma_{i} + \gamma_{j} -\pp{\sqrt{\gamma_{i}} - \sqrt{\gamma_{j}}}^{2}$ \\
    Poisson ratio of particle $i$  		        & $\nu_i$ \\
    \hline
    \end{tabular}
\label{model:table:DEMForces}
\end{table*}

\begin{align}
    \vm{F}_{\mathrm{n},ij} &= k_\mathrm{n} \delta_\mathrm{n} \vm{n}_{ij} + \eta_\mathrm{n} \vm{v}_{\mathrm{n},ij} \label{eq::normal_force}, \\
    \vm{F}_{\mathrm{t},ij} &= \min \pp{ k_\mathrm{t} \vm{\delta}_\mathrm{t} + \eta_\mathrm{t} \vm{v}_{\mathrm{t},ij}, \ \mu_\mathrm{t} \norm{\vm{F}_{\mathrm{n},ij}} \frac{k_\mathrm{t} \vm{\delta}_\mathrm{t} + \eta_\mathrm{t} \vm{v}_{\mathrm{t},ij}}{\norm{k_\mathrm{t} \vm{\delta}_\mathrm{t} + \eta_\mathrm{t} \vm{v}_{\mathrm{t},ij}}}} \label{eq::tangential_force}, \\
    \vm{M}_{\mathrm{t},i} &= - R_i (\vm{F}_{\mathrm{t},ij} \times  \vm{n}_{ij} ). \label{eq::tangential_torque}
\end{align} 

\subsubsection{Cohesive force}
\label{subsec:Cohesive_force}
Metal powders used for additive manufacturing typically have a particle size distribution under \SI{100}{\micro\meter} (Figure~\ref{fig::PSD_SEM_Model_DMT}b). At that scale, cohesive forces due to van der Waals interaction are of the same order of magnitude as gravity \cite{coetzee_2023a,seville_2000}, thus an adequate cohesive force model must be used if we want to accurately simulate the powder spreading process. Two models are mainly used for powder spreading simulations: the Johnson-Kendall-Roberts (JKR) \cite{johnson_1971} and Derjaguin-Muller-Toporov (DMT) \cite{derjaguin_1975}. Here we use a version of the DMT model introduced by Meier et al. \cite{meier_2019} described in equations \eqref{eq::DMT} to \eqref{eq::delta_star}. Although both the JKR and DMT force models are viable \cite{meier_2019}, the DMT model is more appropriate for rigid and small particles \cite{shi_2004}. Moreover, the DMT model does not require solving a non-linear equation for each contact \cite{parteli_2014} which results in a decrease in computational cost compared to the JKR model. The DMT force model used in this study is:

\begin{equation}
        F_{\mathrm{DMT}} =    
        \begin{cases}
            F_{\mathrm{po}}, & \delta_{\mathrm{o}}  \leq \delta_\mathrm{n} \\
            \dfrac{-AR_{\mathrm{e}}}{6 \delta_{n}^2}, & \delta^*  < \delta_\mathrm{n} < \delta_{\mathrm{o}} \\
            0, &  \delta_{\mathrm{n}} \leq \delta^*
        \end{cases}
        \label{eq::DMT}
\end{equation}
with,
\begin{align}
    F_{\mathrm{po}} &= -2\pi\gamma_{\mathrm{e}}R_{\mathrm{e}},\\
    \delta_{\mathrm{o}} &= - \sqrt{\frac{ A R_{\mathrm{e}}}{6 \norm{\vm{F_{\mathrm{po}}}}}} \label{eq::delta_o},\\
    \delta^* &= \frac{\delta_{\mathrm{o}}}{ \sqrt{C_{\mathrm{FPO}}}},
    \label{eq::delta_star}
\end{align}
where $A$ is the Hamaker constant, $\vm{F_{\mathrm{po}}}$ is the pull-off force and $C_{\mathrm{FPO}}$ is a user parameter, set at $0.1$ in this study, which controls the distance at which long-range cohesive forces are considered negligible relative to $\vm{F_{\mathrm{po}}}$. Figure~\ref{fig::PSD_SEM_Model_DMT}d illustrates the norm of the total normal force, assuming no relative velocity between the particles, as a function of the normalized overlap using ($\delta_n / |\delta_o| $). In the range $-3 \leq \delta_n / |\delta_o| \leq -2 $, we see a discontinuity of the total normal force. As explained previously, the overlap at which this discontinuity is located is controlled by $C_{\mathrm{FPO}}$. Between $-1 \leq \delta_n / |\delta_o| \leq 0$, only a constant cohesive force is applied to the particles. When the overlap becomes positive, the elastic portion of the normal contact force begins to increase nonlinearly, and the total normal force progressively becomes repulsive.

\subsubsection{Rolling resistance}
\label{subsec:rolling_res}
Rolling resistance models are used to take into account the non-sphericity of particles \cite{ai_2011} and can greatly affect the powder rheology. Different rolling resistance models have been used to simulate the powder spreading step, from constant rolling friction models \cite{he_2020} to viscous ones \cite{meier_2019,wang_2021, weissbach_2024}, but they are not always explicitly stated in all research article that are considering spherical particles \cite{shaheen_2021, jaggannagari_2025, deng_2025,hu_2025, wang_2020}. Both of these models have their limitation when simulating a static or a dynamic granular problem \cite{ai_2011}, which are simultaneously present during the spreading process (on the building plate between two recoater passes and in the powder heap in front of the recoater). Furthermore, the angle of repose (AoR), which is commonly used to calibrate the surface properties in DEM simulations \cite{meier_2019,wang_2020,yim_2022}, is greatly influenced by the choice of rolling friction model. The elastic-plastic spring-dashpot (EPSD) rolling resistance model can adequately simulate static and dynamic granular systems~\cite{ai_2011}. For this reason, we use this model to simulate the spreading process. To our knowledge, the present work is the first publication related to powder spreading simulations in additive manufacturing that uses this rolling friction model. The EPSD rolling resistance model is given by the following set of equations:
\begin{subequations}
\begin{align}
   \vm{\omega}_{ji} &= \vm{\omega}_{i} - \vm{\omega}_{j}, \\
   \vm{\omega}_{ji,\mathrm{p}} &= \vm{\omega}_{ji} - \pp{\vm{\omega}_{ji}\cdot\vm{n}_{ij}}\vm{n}_{ij},\\
   \vm{\Delta\theta} &= \Delta t \; \vm{\omega}_{ji,\mathrm{p}},\\
   k_\mathrm{r} &= 2.25 k_\mathrm{n} \pp{\muRE \Ree}^2,\\
   \vm{\Delta M}_{\mathrm{r},t} ^\mathrm{k} &= - k_\mathrm{r}\vm{\Delta\theta},\\
   \vm{M}_{\mathrm{r}, t} ^\mathrm{k} &=  \vm{M}_{\mathrm{r},t-\Delta t}^\mathrm{k} + \vm{\Delta M}_{\mathrm{r},t}^\mathrm{k}, \\
   M^\mathrm{m}_\mathrm{r} &= \muRE R_e \norm{\vm{F}_{n,ij}},\\
   \vm{M}^k_{\mathrm{r},t} &= \begin{cases}
         \vm{M}^\mathrm{k}_{\mathrm{r},t}, & \norm{\vm{M}^\mathrm{k}_{\mathrm{r},t}} < M^\mathrm{m}_\mathrm{r} \\
         \frac{\vm{M}^\mathrm{k}_{\mathrm{r},t} }{\norm{\vm{M}^\mathrm{k}_{\mathrm{r},t}}} M^\mathrm{m}_\mathrm{r}, & \mathrm{else}  
    \end{cases},\\
    C_r &= 2 \eta_r \sqrt{I_e K_r},\\
    \vm{M}^d_{r,t} &= \begin{cases}
         -C_r \vm{\omega}_{ij,t} , & \norm{\vm{M}^k_{t}} < M^m_r \\
         -f C_r \vm{\omega}_{ij,t}, & \mathrm{else} 
    \end{cases},\\
    \vm{M}_{\mathrm{r},t} &= \vm{M}^k_{\mathrm{r},t} + \vm{M}^d_{\mathrm{r},t},
\end{align}\label{dem:rol_fric:EPSD}
\end{subequations}
\noindent where $\vm{\omega}_{ji}$ is the relative angular velocity between particle $i$ and $j$, $\vm{\omega}_{ij,p}$ is the relative angular velocity in the contact plane, $\vm{\Delta\theta}$ is the incremental relative rotation, $k_r$ is the rolling stiffness constant, $\vm{\Delta M}_{r,t}$ is the incremental elastic rolling resistance torque, $M^\mathrm{m}_\mathrm{r}$ is the limiting spring torque norm, $C_r$ is the rolling viscous damping constant, $f$ is the full mobilization model parameter which is set to $0.1$ in this study. $\vm{M}^k_r$ and $\vm{M}^d_r$ are the elastic and viscous damping torques, respectively.

\subsection{Simulation set-up}
\label{subsec::simulation_set-up}
The simulation set-up for our powder spreading simulation is similar to the experimental set-up, with every component present, i.e., the feedstock platform, measuring platform, gap, moving spreader, etc. However, modeling strategies are required to minimize the computational cost, while maintaining the adequacy between the experiments and the simulations.

To keep the number of particles manageable throughout all spreading simulations (up to 1.7 millions at any given time), the computational domain is reduced. First, the dept of the domain ($z$ axis) is set to $\SI{2}{\milli\meter}$ ($\approx 20 d_{90}$) with the use of a periodic boundary condition. This is commonly used in powder spreading DEM simulation \cite{jaggannagari_2025, yao_2021}. The sensitivity of the simulation results to the depth of the domain is studied in \ref{app::sensibility_domain_dept}. Next, the length ($x$ axis) of every component, except for the gap and the spreader, are multiplied by a factor $\frac{1}{7}$ (L$_1$), $\frac{2}{7}$ (L$_2$), $\frac{3}{7}$ (L$_3$) or $\frac{4}{7}$ (L$_4$), depending on the simulation,  which gives the measuring plate an approximate length of $\SI{1}{\centi\meter}$, $\SI{2}{\centi\meter}$, $\SI{3}{\centi\meter}$ and $\SI{4}{\centi\meter}$, respectively compared to $\SI{7}{\centi\meter}$ in the real experiment. Particles falling into the gap or not settling on the measuring plate are removed from the simulation when they cross the $y_\mathrm{min}$ and $x_\mathrm{max}$ boundaries of the simulation domain.

Moreover, some modeling limitations force us to change $\delta_\mathrm{s}$ to \SI{100}{\micro\meter}. Keeping  $\delta_\mathrm{s}\approx \SI{0}{\micro\meter}$ results in excessive overlap of the particles when the spreader is over the transfer plate. Particles are jamming between the spreader and the transfer plate, resulting in a non-physical behavior. 

\subsection{DEM calibration and powder physical properties}
\label{subsec::dem_calib_and_PS}
To find appropriate values for the sliding friction (${\muT}_{,i}$), rolling friction (${\muR}_{,i}$) and surface energy ($\gamma_i$) for the powder, a calibration experiment is performed. We use a commercially available experimental device called a Granuheap\texttrademark \cite{shi_2020} manufactured by Granutools, which builds powder heaps in a repeatable manner by lifting a cylinder filled with powder. Our experiment consists in using \SI{5}{\gram} of powder, ($\approx 1.6$ million particles), inside a $\SI{1}{\centi\meter}$ inner diameter cylinder. The lifting velocity of the cylinder is set to $\SI{5}{\centi\meter\per\second}$ and the experiment is repeated 20 times. In each experiment, a camera captures 16 distinct photographs of the powder heap profile in different orientations. Then, an average experimental static angle of repose (S-AoR) and an average heap profile between the 20 experiments are computed using an in-house image analysis code. These two experimental measurements are then compared to calibration DEM simulations. 

We then perform one-to-one simulations of the Granuheap\texttrademark using \lethe{} while varying $\muT{}_{,i}$, $\muR{}_{i}$ and $\gamma_i$ to find the best fit between experimental and numerical S-AoR and heap profiles. The restitution coefficient (${e}_i$), rolling viscous damping coefficient ($\eta_\mathrm{r}$) and full mobilization model parameter ($f$) are set to constant values of $0.9$, $0.3$ and $0.1$, respectively, for every simulation as the spreading process is believed to be less sensible to them. 

In this study, three different property sets (PS) of DEM surface properties are used for our powder spreading simulations. Every PS (PS1, PS2 and PS3) obtained similar S-AoR and heap profiles during their respective calibration simulation while having distinctive DEM model parameters. Table \ref{tab::calibration::dem_param} shows these sets of parameters as well as their respective S-AoR values. Finally, Table \ref{tab::powder_physical_properties} presents the powder physical properties in use in every simulation of this study. Lowering the Young's modulus of the particles to achieve a manageable time step ($\Delta t$) is common practice in DEM simulation \cite{chen_2017}. Here, it is reduced by a factor of 4000 compared to the real value to achieve a $\Delta t = \SI{2.816e-7}{\second}$, equivalent to $15\%$ of the Rayleigh time of the smallest particle in the simulation, which is commonly being used \cite{blais_2019}. The properties of the walls are fixed equally to the properties of the particles as a sake of simplicity.

\begin{table*}[!hbt]
\caption{DEM model parameter used and S-AoR obtained in calibration simulation for every property set (PS). The experimental S-AoR was $\SI{37.5}{\degree}$}
\centering
\begin{tabular}{c | c  c  c}
    DEM model parameter & PS1 & PS2 & PS3 \\   
    \hline
    Slidng friction $(\muT{}_{,i})[-]$    &  0.2 & 0.2 & 0.4 \\
    Rolling friction $(\muR{}_{,i})[-]$    &  0.1 & 0.2 & 0.35\\
    Surface energy $(\gamma_i) [\si{\si{\joule\per\square\meter}}]$ & 35e-5 & 25e-5 & 8e-5 \\
    Restitution coefficient $({e}_i)[-]$    &  0.9  &  0.9  & 0.9  \\
    Rolling viscous coefficient $(\eta_\mathrm{r})[-]$   &  0.3  &  0.3  & 0.3  \\
    $f[-]$                 &  0.1  &  0.1  & 0.1  \\
    Static angle of repose (S-AoR)$[\si{\degree}]$   & 35.79 & 36.34 & 37.18\\
    \end{tabular}
    \label{tab::calibration::dem_param}
\end{table*}

\begin{table}[!htb]
\caption{Powder physical properties used in the numerical simulations.}
\centering
\begin{tabular}{c | c}
    Physical properties & Numerical value \\   
    \hline
    Poisson's ratio ($\nu$) [$-$]   & 0.342\\
    Young's modulus ($Y$) [$\si{\mega\pascal}$] & 26.25\\
    True density ($\rho$) [$\si{\kilo\gram\per\meter\cubed}$] & 4386 \\
    Hamaker constant ($A$) [\si{\joule}] & $4\times10^{-19}$ \\
\end{tabular}
\label{tab::powder_physical_properties}
\end{table}

\section{Results} 
\label{sec::results}
\subsection{Experimental measurements} 
\label{subsec::exp_meas_den_phe}
\begin{figure}[!ht]
        \centering
        \includegraphics[width=\linewidth]{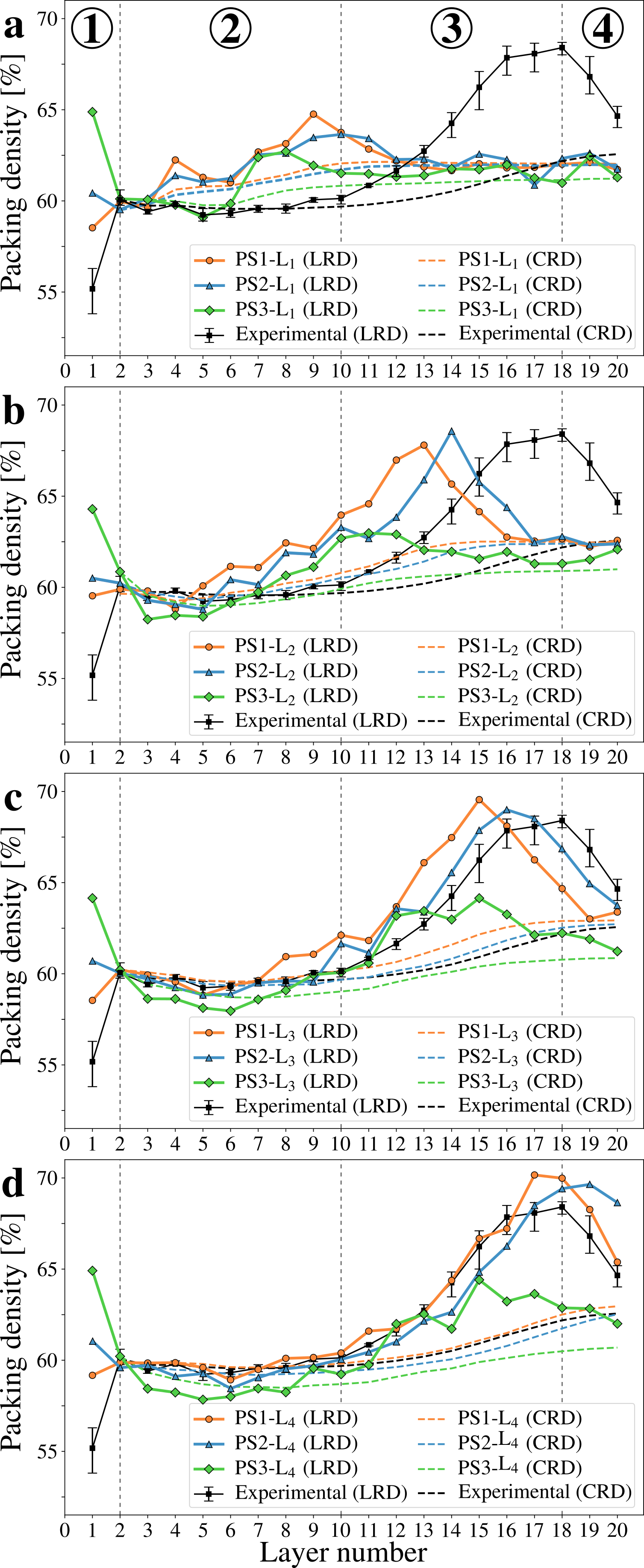}
        \caption{\textbf{Experimental and numerical results comparison.} Cumulative relative density (\CRD{}) and layer effective relative density (\LRD{}) for every property set (PS1, PS2 and PS3) and domain length (L) combinaisons. Results for the domain length a) L$_1$, b) L$_2$, c) L$_3$ and d) L$_4$. When the domain length is increased, the \CRD{} and \LRD{} values of PS1 and PS2 approach the experimental measurements, while PS3 remains largely unchanged.}
        \label{fig::LRD_and_CRD}
\end{figure}

Figure~\ref{fig::LRD_and_CRD}a presents the average \LRD{}, shown as a solid black curve, and \CRD{}, shown as a dashed black curve, measured across the three repeated experiments as a function of the layer number. These curves are also reproduced in Figure~\ref{fig::LRD_and_CRD}b, \ref{fig::LRD_and_CRD}c and \ref{fig::LRD_and_CRD}d. The error bars indicate the minimum and maximum values observed across all experiments, which were each repeated three times. Four zones characterize the relative density plots. The first zone (1, first layer effect), where there is a large discrepancy between \LRD{}$_1$ and \LRD{}$_2$. This discrepancy can be explained by the total thickness of the powder bed, equal to \SI{300}{\micro\meter} at layer 2, which is close to $3d_{90} = \SI{318}{\micro\meter}$. This is consistent with the work of Yao et al.~\cite{yao_2021}, where the relative density of single layers significantly increased when reaching a layer thickness of $3d_{90}$. The second zone (2, plateau), where \LRD{} is constant from layer 2 to 10, a third zone (3, densification), where \LRD{} increases from layer 10 to 18, and a fourth zone (4, decrease) where \LRD{} decreases from layer 18 to 20. 

\subsection{Simulation results}\label{sec::num_resuls}
\begin{figure}[!htb]
        \centering
        \includegraphics[width=\linewidth]{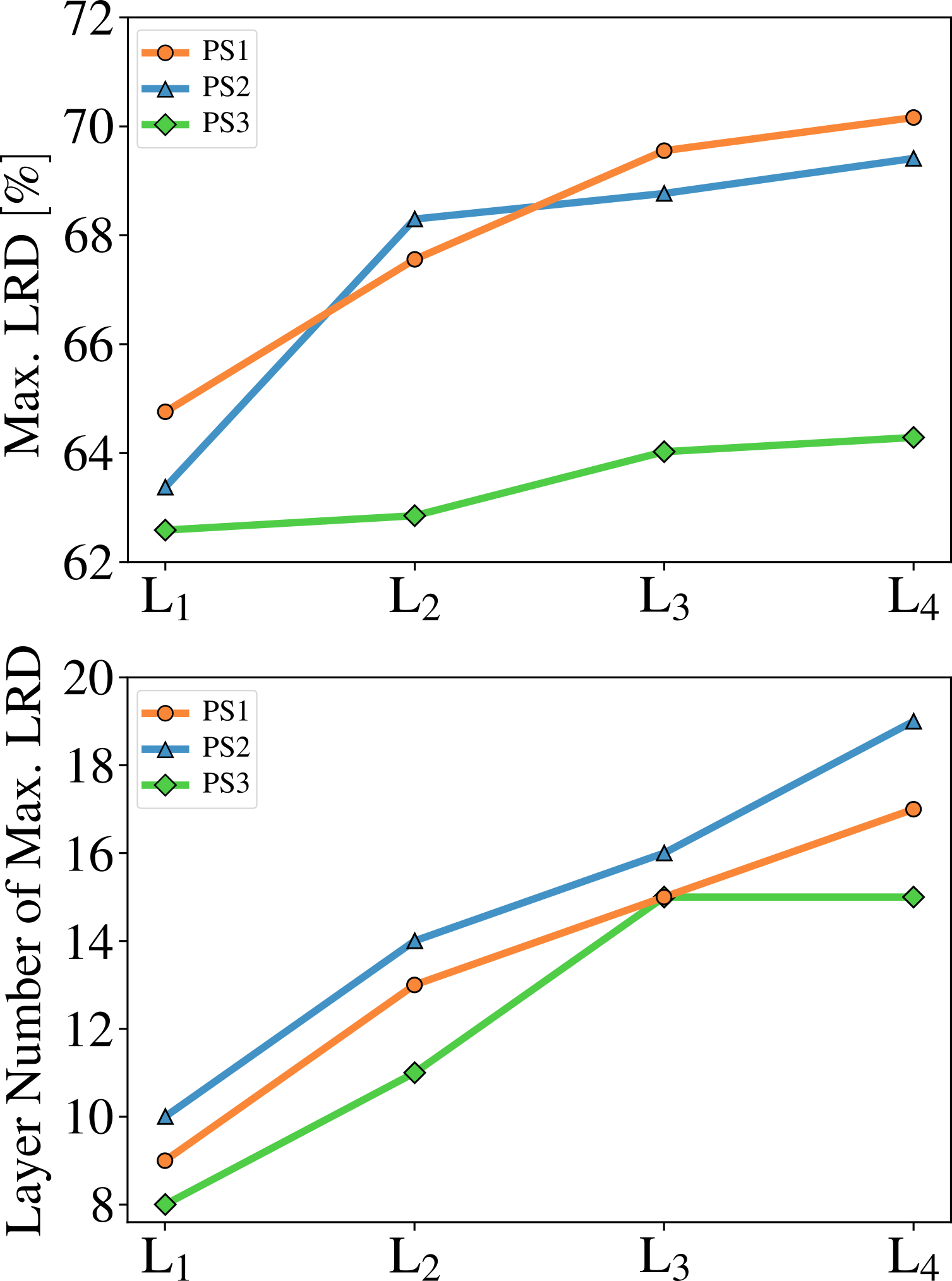}
        \caption{\textbf{Overview of the densification phenomenon.} Maximum effective layer relative density (LRD) reached during spreading simulation using each property set (PS) as a function of the domain length (L)  (Top). Layer number at which this maximum LRD is reached as a function of the domain (Bottom).}
        \label{fig::amplitude_and_layer}
\end{figure}

The powder spreading simulations are performed for every PS and L combinations (see Movie S-PS1-L1, S-PS3-L1, S-PS1-L4 and S-PS3-L4). Figures~\ref{fig::LRD_and_CRD}a–d present the LRD (solid curves) and CRD (dashed curves) results as a function of the layer number for domain lengths L$_1$, L$_2$, L$_3$, and L$_4$, respectively. We can see that the agreement with the experimental results improves as the domain length increases. Figure~\ref{fig::amplitude_and_layer} shows the maximum \LRD{} value reached between \LRD{2} to \LRD{20} and the layer number at which this maximum value is reached, both as a function of the domain length.  Although densification occurs for all domain length, the amplitude of the densification as well of the layer number at which it occurs increases with the domain length. For longer domain length (L$_3$ and L$_4$), the DEM model with PS1 or PS2 set of physical properties reproduces the experimental results without any form of additional calibration or tuning. Consequently, we can be confident that the DEM model can provide additional insight into the physics of the densification mechanism, which we obtain by analyzing the PS1-L$_2$ and PS3-L$_2$ simulations in the following sections. The same observations can be drawn from the L$_3$ and L$_4$ domain lengths, but they are more readily highlighted with the L$_2$ due to the powder bed thickness ($y$) to length ($x$) ratio. 

\subsubsection{Layer-Based Particle Analysis}
To gain insight of the underlying physical mechanism causing the densification phenomenon reproduced during our DEM simulation, we analyze the powder bed by coloring the particles according to their layer number during the spreading process (see Movie LA-PS1-L2 and LA-PS3-L2).

\begin{figure*}[!htb]
        \centering
        \includegraphics[width=\linewidth]{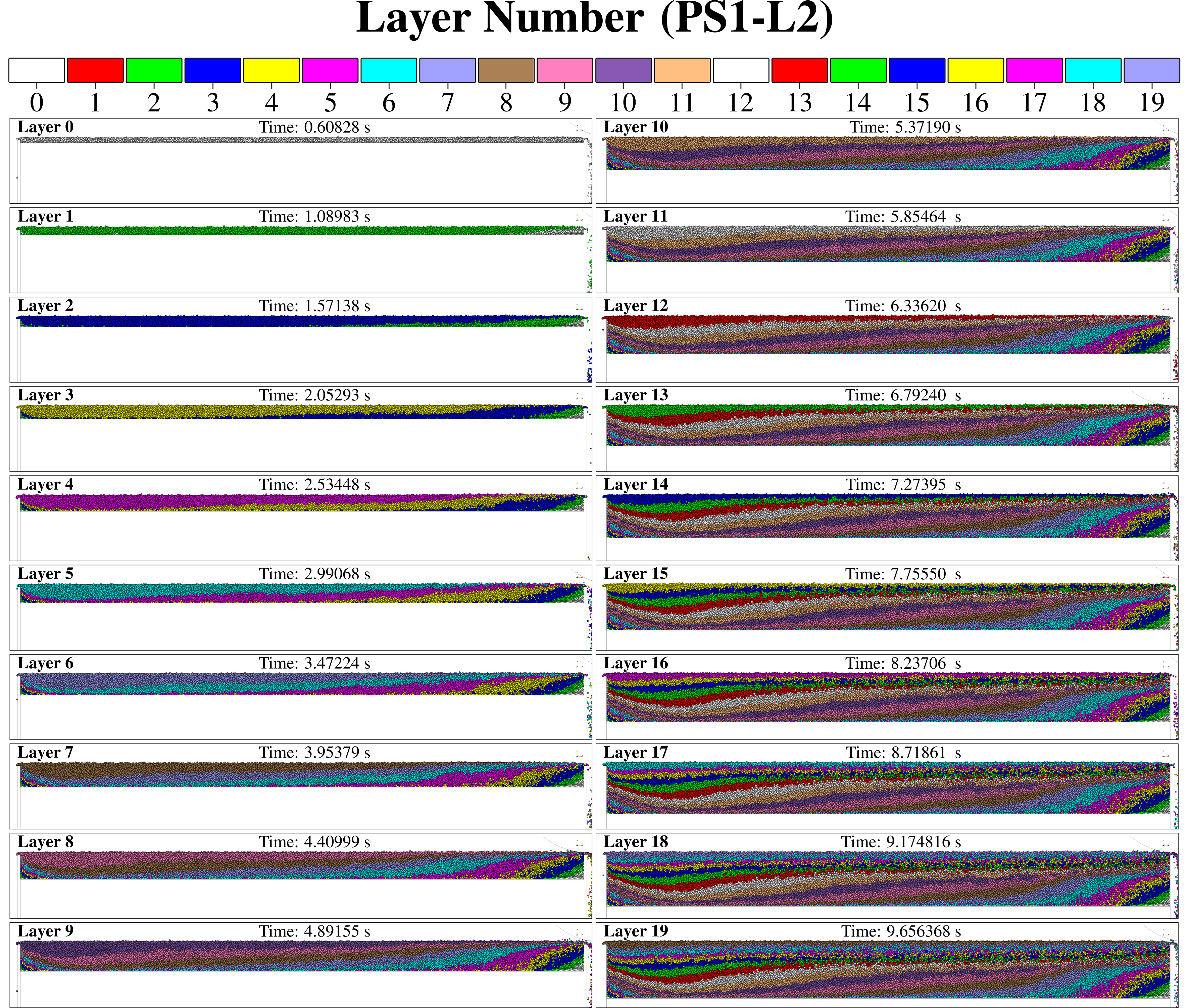}
        \caption{\textbf{Colored particles according to their layer number during the spreading process for the PS1-L$_2$ simulation.} Visualization of the powder scraping using particle coloring relative to their layer number from layer 0 to layer 19 for the PS1-L$_2$ simulation. }
        \label{fig::particle_removal_colors_PS1-L2}
\end{figure*}

Figure~\ref{fig::particle_removal_colors_PS1-L2} displays a side view ($x$-$y$ plane) of the measuring plate after each spread layer for the PS1-L$_2$ simulation, where significant densification occurs. Two main elements can be observed. First, in layer 1 to 8, the newly spread layers are removing a significant amount of particles from previous layers and penetrate deeply into the powder bed at the onset (left side) of the measuring plate. As the spreader moves forward, the layers gradually thin out along the length of the plate. During this stage, considerable particle displacement is observed at the bottom of the powder bed, where particles slide toward the end (right side) of the plate and are eventually removed from it. From layer 14 to 19, the newly spread layers have a more uniform thickness along the length of the measuring plate, less particle displacement is observed at the bottom of the powder bed and static zones are visible at the beginning and end of the plate. The transition between these two behaviors occurs between Layer 9 to 13,  where the \LRD{} is increasing and peaking. During this transition, the middle of the powder bed becomes more static after each layer.

\begin{figure*}[!htb]
        \centering
        \includegraphics[width=\linewidth]{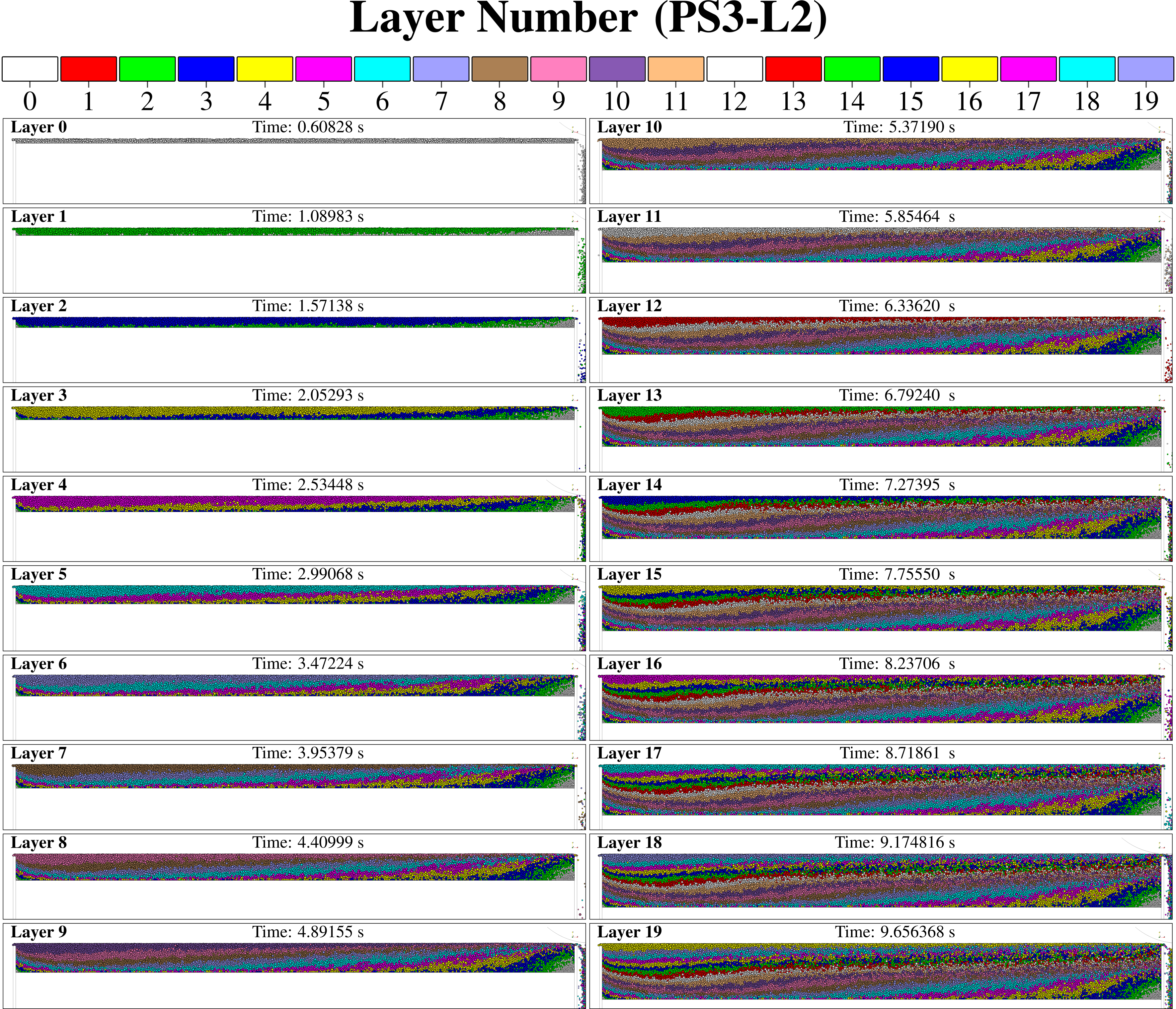}
        \caption{\textbf{Colored particles according to their layer number during the spreading process for the PS3-L$_2$ simulation.} Visualization of the powder scraping using particle coloring relative to their layer number from layer 0 to layer 19 for the PS3-L$_2$ simulation. }
        \label{fig::particle_removal_colors_PS3-L2}
\end{figure*}

Figure~\ref{fig::particle_removal_colors_PS3-L2} presents the same side views as Figure \ref{fig::particle_removal_colors_PS1-L2}, but for PS3-L$_2$, which does not exhibit the same degree of densification. Notable differences can be observed between the two figures. First, the newly spread layers in PS3-L$_2$ do not penetrate as deeply into the powder bed. Second, previously spread particles do not slide as readily toward the end of the measuring plate, compared to PS1-L$_2$. This can be easily observed with yellow particles from layer 3, which are still visible at the left of the plate after the spreading of layer 4 for the PS3-L$_2$ but not for the PS1-L$_2$. This is coherent with the higher $\muT{}$ coefficient of the PS3.

We analyze the displacement field (DF) of particles between subsequent layers. The DF of layer $k$ (DF$_k$) illustrates the particle displacement in the $x$-$y$  plane. These particles were located on the measuring plate after the spreading of layer $k-1$, and their movement is recorded during the spreading of layer $k$ and displayed in the associated DF. Each vector originates at a particle's location after layer $k-1$ and is oriented in the direction of the particle displacement that occurred during layer $k$. For visualization purposes, vector lengths are scaled by a factor $\frac{1}{10}$. Particles that are close to stationary relative to the measuring plate during the spreading of layer $k$ are represented by dots. Finally, particles removed from the measuring plate during the spreading of layer $k$ are not represented, resulting in white space on the DF. This tool allows us to visualize which regions of the powder bed are static.

\begin{figure*}[!htb]
        \centering
        \includegraphics[width=\linewidth]{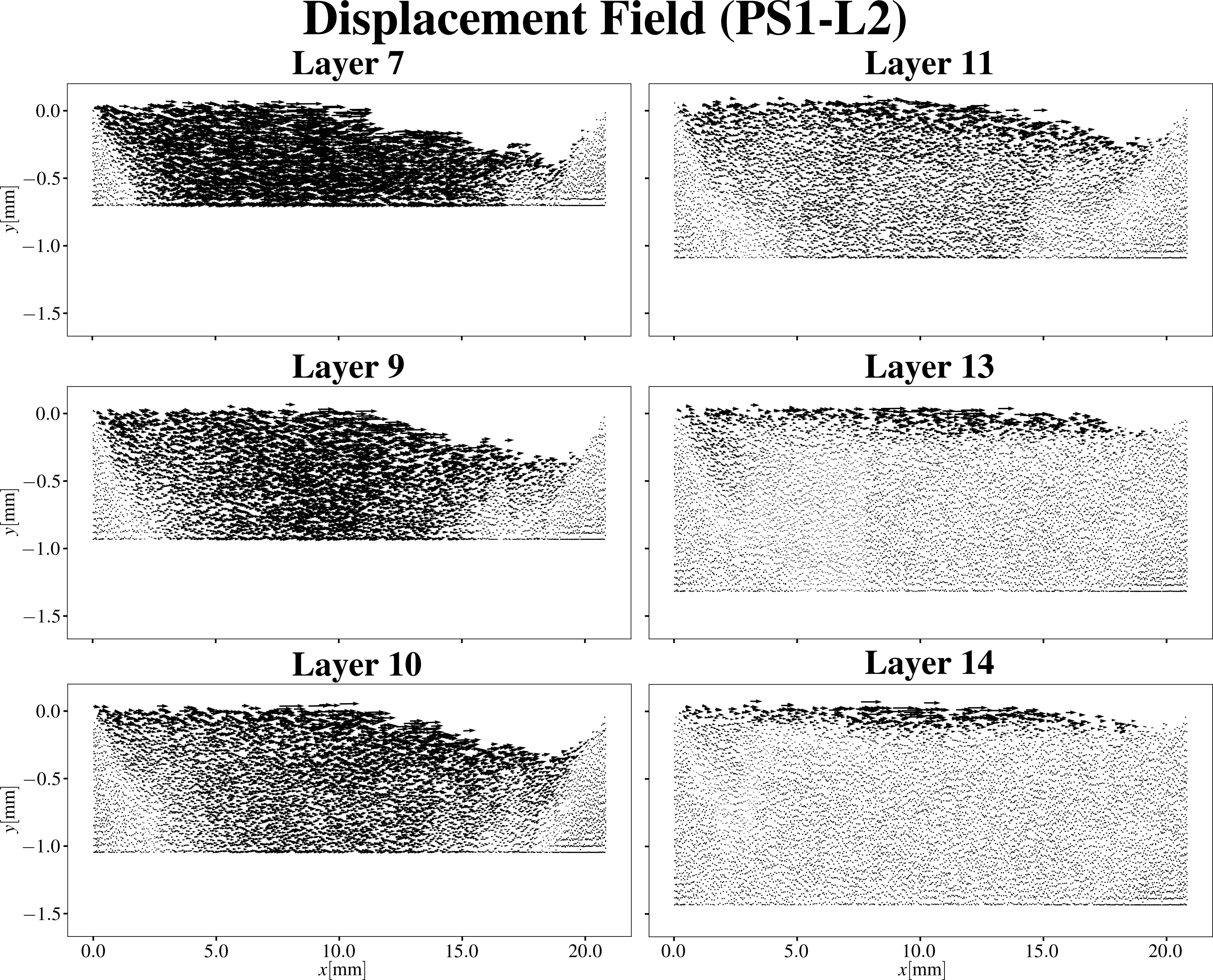}
         \caption{\textbf{Displacement field PS1-L$_2$.} Displacement of the particle remaining on the measuring plate after layer 6, 7, 8, 9, 11 and 12 for the PS3-L$_2$ simulation. Vectors' lengths are scaled by a factor $\frac{1}{10}$. Dots represent particles that didn't move significantly relative to the measuring plate.}
        \label{fig::displacement_field_PS1-L2}
\end{figure*}
Figure~\ref{fig::displacement_field_PS1-L2} illustrates the DFs of the PS1-L$_2$ simulation for the layer 7, 9, 10, 11, 13 and 14 (See supplementary material DF-PS1-L2, DF-PS2-L2 and DF-PS3-L2  to see all the DFs). In the DF$_7$, we observe a large particle removal (indicated by a large white zone) near the end of the measuring plate, resulting from the particle scraping as shown in Figure~\ref{fig::particle_removal_colors_PS1-L2}. Additionally, two static zones begin to form at the start and end of the measuring plate, caused by the interaction between the spreader and the vertical walls surrounding the measuring plate. These static zones grow toward the middle as the number of layers increases, as can be seen in the associated DF$_9$ and DF$_{10}$. After layer 10, when the \LRD{} starts to rise substantially, the static zones are close enough to restrict the displacement of particles at the center of the plate. This restriction allows the spreader to compact the particles during the spreading of layers 10 to 13, instead of scraping them off. By layer 13, the majority of the particles are compressed in the middle, peak \LRD{} is reached, and no particle movement can be observed at the bottom of the powder bed. No clear difference can be seen between layer 13 and the following layers. When the domain length increases, the initial distance between the two static zones is greater, and consequently, their merge occurs for a greater number of layers. Thus, the densification process begins with a thicker powder bed, which provides more real estate for compaction, resulting in higher \LRD{} values. This explains why increasing the domain length changes the amplitude of the densification and the layer number at which it is peaking.  

\begin{figure*}[!htb]
        \centering
        \includegraphics[width=\linewidth]{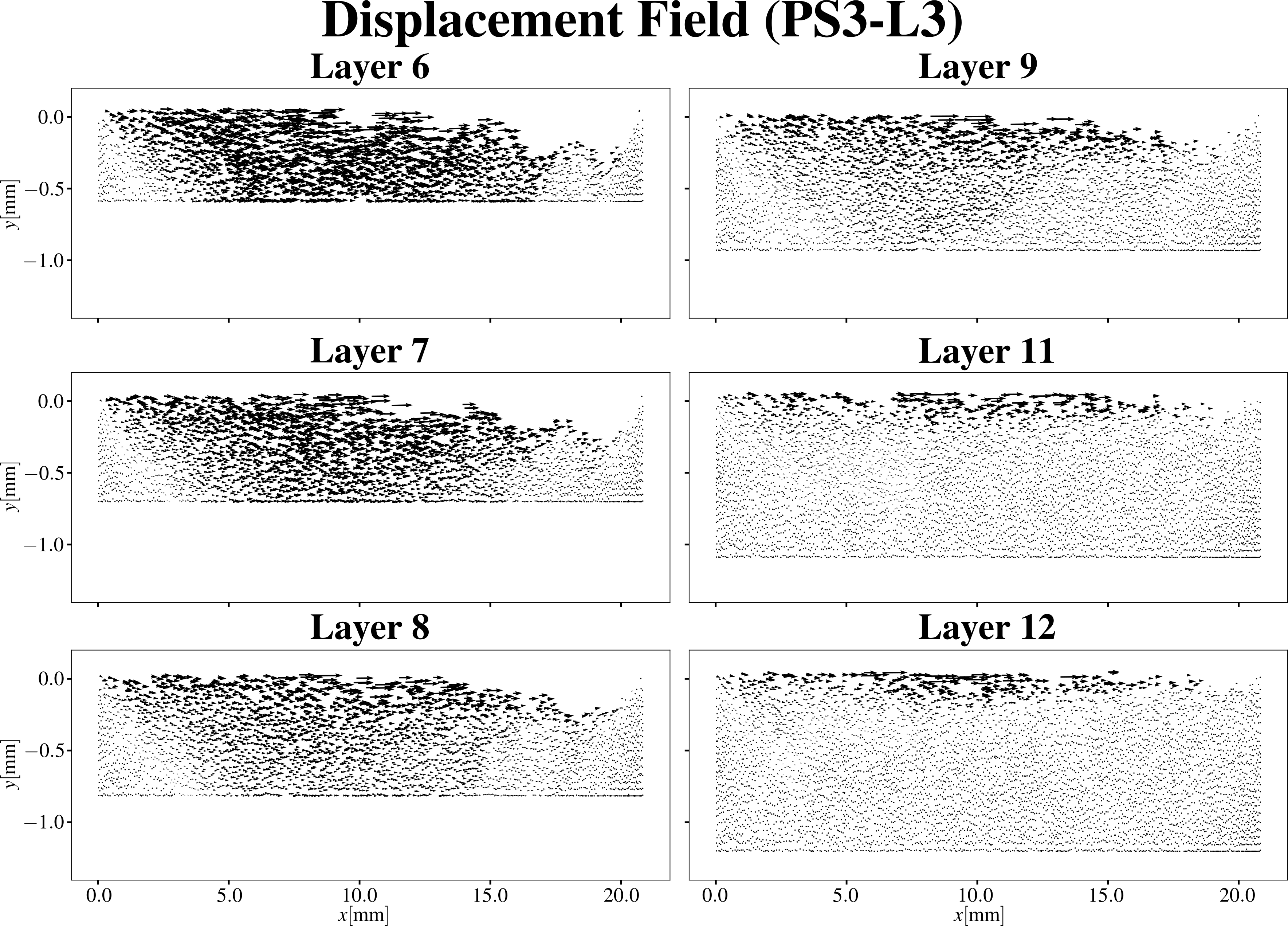}
        \caption{\textbf{Displacement field PS3-L$_2$.} Displacement of the particle remaining on the measuring plate after layer 6, 7, 8, 9, 11 and 12 for the PS3-L$_2$ simulation. Vectors' lengths are scaled by a factor $\frac{1}{10}$. Dots represent particles that didn't move significantly relative to the measuring plate.}
        \label{fig::displacement_field_PS3-L2}
\end{figure*}

Looking at the DFs of the PS3-L$_2$ simulation, in Figure~\ref{fig::displacement_field_PS3-L2}, we observe the same two behaviors explained in Figure~\ref{fig::displacement_field_PS1-L2} but at lower layer number. At layer number 7 and prior, particles are being scraped away from the build plate. Then at layer 8, when the \LRD{} starts to rise (Figure~\ref{fig::LRD_and_CRD}b), the two static zones are merging. Finally, no particle movement at the bottom of the powder bed is visible at layer 11 and onward. This tells us that the particle surface properties also influence the densification phenomena and that a simple calibration based on the S-AoR can lead to a combination of DEM model parameters that do not correspond to the metal powder being calibrated and an incorrect densification.  

\FloatBarrier

\section{Conclusion} 

In this study, powder spreading experiments using a novel experimental set-up were conducted highlighting a previously unreported densification phenomenon occurring at higher layer numbers. Large-scale DEM simulations were carried, using three different sets of DEM model parameters, calibrated using a static angle repose experiment, to replicate this densification phenomenon and explain its causes. The following conclusions can be drawn :
\begin{enumerate}
    \item The densification phenomenon is related to the length of the measuring (or building) plate in the spreading direction. Densification starts occurring when the front and back static zones, created by the vertical walls at the edges of the measuring plate, join in the middle of that measuring plate. This which prevents the particles from sliding when the recoater is laying a new layer. This suggests that vertical walls formed by the part being built during the manufacturing process can create these static zones, which could result in a localized densification in the powder bed. 
    \item A significant amount of powder previously deposited on the measuring plate is removed from the powder bed by the recoater at lower layer numbers. This suggests that metal particles that have undergone one or many complete thermal cycles are being removed from the building plate during the real manufacturing process, which could aggravate the quality of the powder when it is recycled. Moreover, less powder could potentially be used for the same build if the amount of powder removed is reduced.
    \item Powder surface properties, represented by the DEM model parameters used in this study, affect the amplitude and the layer number at which the densification phenomenon occurs. These surface properties could be tuned, along with the powder spreading process parameters, to achieve layer packing density as constant and as high as possible regardless of the layer number. 
\end{enumerate}
In our future work, the cooperative effect of different process parameters and recoaters, including flexible recoaters, will be studied with our now validated DEM framework.

\section{Acknowledgements}

 The authors acknowledge technical support and computing time provided by the Digital Research Alliance of Canada and Calcul Québec. The authors would like to thank Prof. Jean-Philippe Harvey and Semplor for providing the SEM images of the powder. BB and DM acknowledges financial support from the National Research Council (NRC) Canada through the Collaborative R\&D Initiative grant AM-137-1. BB acknowledges the funding from the Multiphysics Multiphase Intensification Automatization Workbench (MMIAOW) Canadian Research Chair Level 2 in computer-assisted design and scale-up of alternative energy vectors for sustainable chemical processes (CRC-2022-00340).


\appendix

\section{Result sensitivity to the domain dept} \label{app::sensibility_domain_dept}
\begin{figure}[H]
    \centering
    \includegraphics[width=\linewidth]{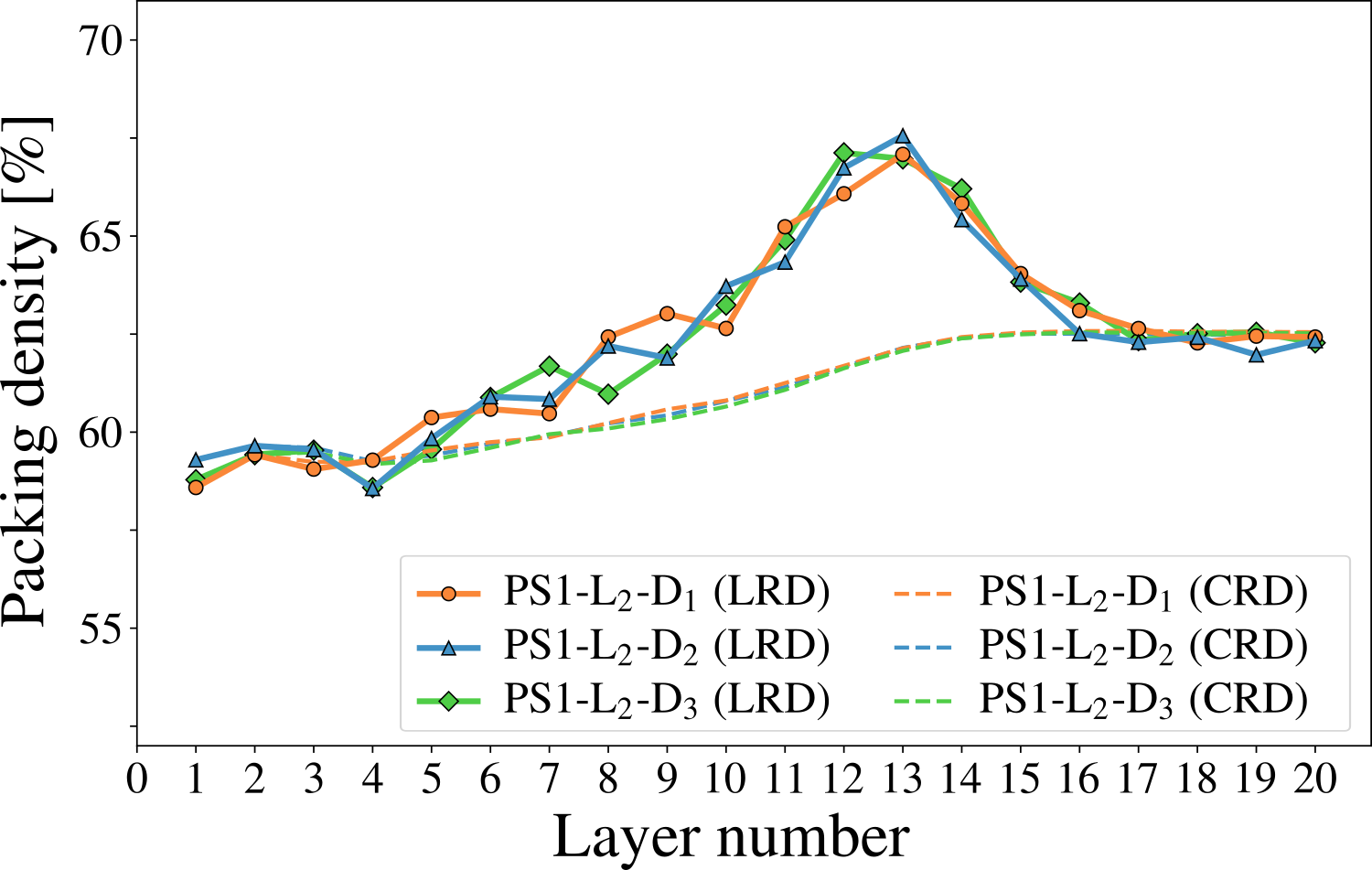}
    \caption{\textbf{Domain depth sensitivity.} LRD and CRD values obtained with the PS1-L$_2$ simulation using three different domain dept (D$_1$,D$_2$ and D$_4$). Associated domain depth values are listed in Table \ref{tab::domain_dept}. }
    \label{fig:domain_depth}
\end{figure}

When using periodic boundary conditions (PBCs), an insufficient domain depth can introduce artificial boundary effects. These include significant interactions between particles and their periodic images, as well as constraints on realistic force propagation within the granular system. Therefore, it is necessary to assess the sensitivity of the simulation results to the domain depth used in this study ($\SI{2}{\milli\meter}$). To do so, the PS1-L$_2$ simulation, which is showing significant densification (Figure~\ref{fig::LRD_and_CRD}) is launched three times using different domain dept values (D$_1$,D$_2$ and D$_4$). These domain dept values are listed in Table~\ref{tab::domain_dept}. Figure~\ref{fig:domain_depth} present the LRD and CRD of those three simulations, where no significant differences is observed. Small differences between the LRD curve are explained by the chaotic behavior of DEM. All three simulation can reproduce the densification. The D$_2$ dept, which is about 20 times the $d_{90}$ of the simulated powder, represent a balance between computational cost and results quality, which is why it has been used throughout this study.

\begin{table}[!htpb]
\caption{Domain depth values used to test the sensitivity.}
\centering
\begin{tabular}{c | c}
    Domain depth & Value \\   
    \hline
    D$_1$ & $\SI{1}{\milli\meter}$ \\
    D$_2$ & $\SI{2}{\milli\meter}$\\
    D$_3$ & $\SI{4}{\milli\meter}$ \\
\end{tabular}
\label{tab::domain_dept}
\end{table}

\FloatBarrier

\section*{Data availability}
Data will be made available on request.

\bibliographystyle{elsarticle-num} 
\bibliography{bibliographie}
\label{app1}

\end{document}